\documentclass[onecolumn,prl,aps,showpacs,amsmath,amssymb,longbibliography]{revtex4-1}

\usepackage{graphicx}
\usepackage{dcolumn}
\usepackage{bm}

\renewcommand{\vec}[1]{{\bf#1}}

\begin{document}

\title{Orthogonal Cherenkov sound in spin-orbit coupled systems}

\author{Sergey Smirnov}
\affiliation{Institut f\"ur Theoretische Physik, Universit\"at Regensburg,
  D-93040 Regensburg, Germany}

\date{\today}

\begin{abstract}
Conventionally the Cherenkov sound is governed by {\it orbital} degrees of
freedom and is excited by {\it supersonic} particles. Additionally, it usually
has a {\it forward} nature with a conic geometry known as the Cherenkov cone
whose axis is oriented {\it along} the {\it supersonic} particle motion. Here
we predict Cherenkov sound of a unique nature entirely resulting from the
electronic {\it spin} degree of freedom and demonstrate a fundamentally
distinct Cherenkov effect originating from essentially {\it subsonic}
electrons in two-dimensional gases with both Bychkov-Rashba and Dresselhaus
spin-orbit interactions. Specifically, we show that the axis of the
conventional {\it forward} Cherenkov cone gets a nontrivial {\it quarter-turn}
and at the same time the sound distribution strongly localizes around this
rotated axis being now {\it orthogonal} to the {\it subsonic} particle
motion. Apart from its fundamentally appealing nature, the orthogonal
Cherenkov sound could have applications in planar semiconductor technology
combining spin and acoustic phenomena to develop, {\it e.g.}, acoustic
amplifiers or sound sources with a flexible spin dependent orientation of the
sound propagation.
\end{abstract}


\maketitle
Dating from the idea of a spin transistor \cite{Datta}, systems with
spin-orbit interactions have been attracting vasty interest because of a
possibility to access the electronic spin degree of freedom by exclusively
electric means and currently constitute a considerable part of contemporary
spintronics \cite{Zutic,Fabian}.

The interaction between the orbital and spin degrees of freedom is a
relativistic effect and qualitatively it may be understood as a transformation
of electric fields into magnetic fields in the rest system of an electron.

In two-dimensional semiconductor heterostructures two types of spin-orbit
interaction are of particular importance. The first one is the Dresselhaus
\cite{Dresselhaus} spin-orbit interaction due to the inversion asymmetry of
the semiconductor crystal structure. The second one is the Bychkov-Rashba
\cite{Bychkov} spin-orbit interaction appearing in asymmetric structures such
as, {\it e.g.}, asymmetric quantum wells. In realistic systems both of these
spin-orbit interactions are usually present and of comparable strengths which
can be measured through, {\it e.g.}, Shubnikov-de Haas oscillations
\cite{Luo,Schaepers}, photocurrents \cite{Ganichev}, optical monitoring of the
spin precession \cite{Meier}.

A qualitatively different class of condensed matter systems, where spin-orbit
interactions are crucial, is the one of topological insulators
\cite{Hasan_2010,Qi_2011}, the phase of matter where metallic edges coexist
with an insulating bulk. The physics of the metallic edges is governed by
helical states \cite{Wu_2006,Xu_2006}. The surface helical states form Kramers
pairs and the time reversal invariance leads to zero gap (or metallic) nature
of these states while the states in the bulk are of finite gap (or insulating)
character. The surface helical states are intimately linked to the bulk states
and the time reversal invariance constrains the number of the Dirac points by
even numbers.

Spin-orbit interaction mechanisms give rise to fascinating physical phenomena
such as the intrinsic spin-Hall effect \cite{Murakami,Sinova} in
semiconductors or the quantum spin-Hall effect \cite{Roth_2009} in topological
insulators, persistent spin helix \cite{Koralek} as an interplay between the
Bychkov-Rashba and Dresselhaus mechanisms in semiconductors or spin-transfer
torques \cite{Mellnik_2014} in topological insulators with applications to
non-volatile memory.

The examples above spin around the electron dynamics. The spin-orbit physics
has, however, another side related to the dynamics of the crystal lattice or
phonons. Although the electron-phonon interaction has an orbital nature, the
dynamics of the electronic orbital degrees of freedom is strongly affected by
the electron spin degree of freedom in systems with spin-orbit
interactions. Therefore, the impact of the electron spin on the phonon
dynamics is of fundamental and practical interest. While the impact of phonons
on the electron spin dynamics was extensively studied ({\it e.g.},
Dyakonov-Perel' \cite{Dyakonov} mechanism or spin decay in quantum dots
\cite{Golovach,Sherman}), the role of the electron spin in the phonon dynamics
was less explored.

One important aspect of the phonon dynamics is the acoustic Cherenkov effect,
a counterpart of the optical Cherenkov effect \cite{Cherenkov,Tamm} where a
medium emits a forward light cone under the action of superlight particles
passing through this medium with velocities larger than the medium speed of
light. Likewise, a medium emits a forward sound cone excited by supersonic
particles whose velocity exceeds the medium sound velocity.

The presence of spin-orbit interaction mechanisms changes this picture. In
two-dimensional semiconductor heterostructures with the Bychkov-Rashba
spin-orbit interaction supersonic particles excite sound filling the forward
Cherenkov cone and also the outward Cherenkov cone containing backward or
anomalous (optical anomalous Cherenkov effect \cite{Joannopoulos} exists too
but, in contrast to the present case, due to the spatial inhomogeneity)
Cherenkov sound \cite{Smirnov_2011}. In other words, the Cherenkov cone angle
which is, conventionally, between 0 and $\pi/2$ extends to $\pi$. On surfaces
of three-dimensional topological insulators the Cherenkov sound is excited by
helical particles which are always supersonic since the Dirac velocity is well
above the sound velocity. In this case the geometry of the Cherenkov sound is
also conic with the Cherenkov cone angle exceeding $\pi/2$, {\it i.e.}, the
anomalous Cherenkov sound is generated in this case too. What makes the
Cherenkov sound in topological insulators distinct is that at high energies it
may strongly localize along certain directions
\cite{Smirnov_2013}. Additionally, at low energies a magnetic field control of
the Cherenkov sound in topological insulators may be of practical interest
\cite{Smirnov_2014}.

The arc of vision above might suggests that spin-orbit interaction mechanisms
enlarge the Cherenkov cone angle but, nevertheless, the very core of the
Cherenkov physics remains unchanged: 1) the Cherenkov sound is still excited
by supersonic particles; 2) its geometry still represents a single cone
(although, the cone angle may exceed $\pi/2$); 3) the cone axis is still
oriented along the direction of motion of the supersonic particle exciting the
Cherenkov sound.

In the following we explore the acoustic Cherenkov effect in realistic
two-dimensional semiconductor heterostructures with both Bychkov-Rashba and
Dresselhaus spin-orbit interaction mechanisms of comparable strengths and
demonstrate that peculiar coupling between the orbital and spin dynamics
results in an acoustic Cherenkov effect of a unique nature fundamentally
different from what has been known so far: 1) the Cherenkov sound is generated
by essentially subsonic electrons; 2) the geometry of the Cherenkov sound
excited by subsonic electrons represents a double cone; 3) the axis of the
Cherenkov double cone gets a quarter-turn and, therefore, is orthogonal to the
direction of motion of the subsonic particle exciting the Cherenkov sound; 4)
the Cherenkov sound distributed within this rotated Cherenkov double cone is
strongly localized around the cone axis or, in other words, the Cherenkov
sound acquires an orthogonal nature.
\begin{figure}
\includegraphics[width=14.0 cm]{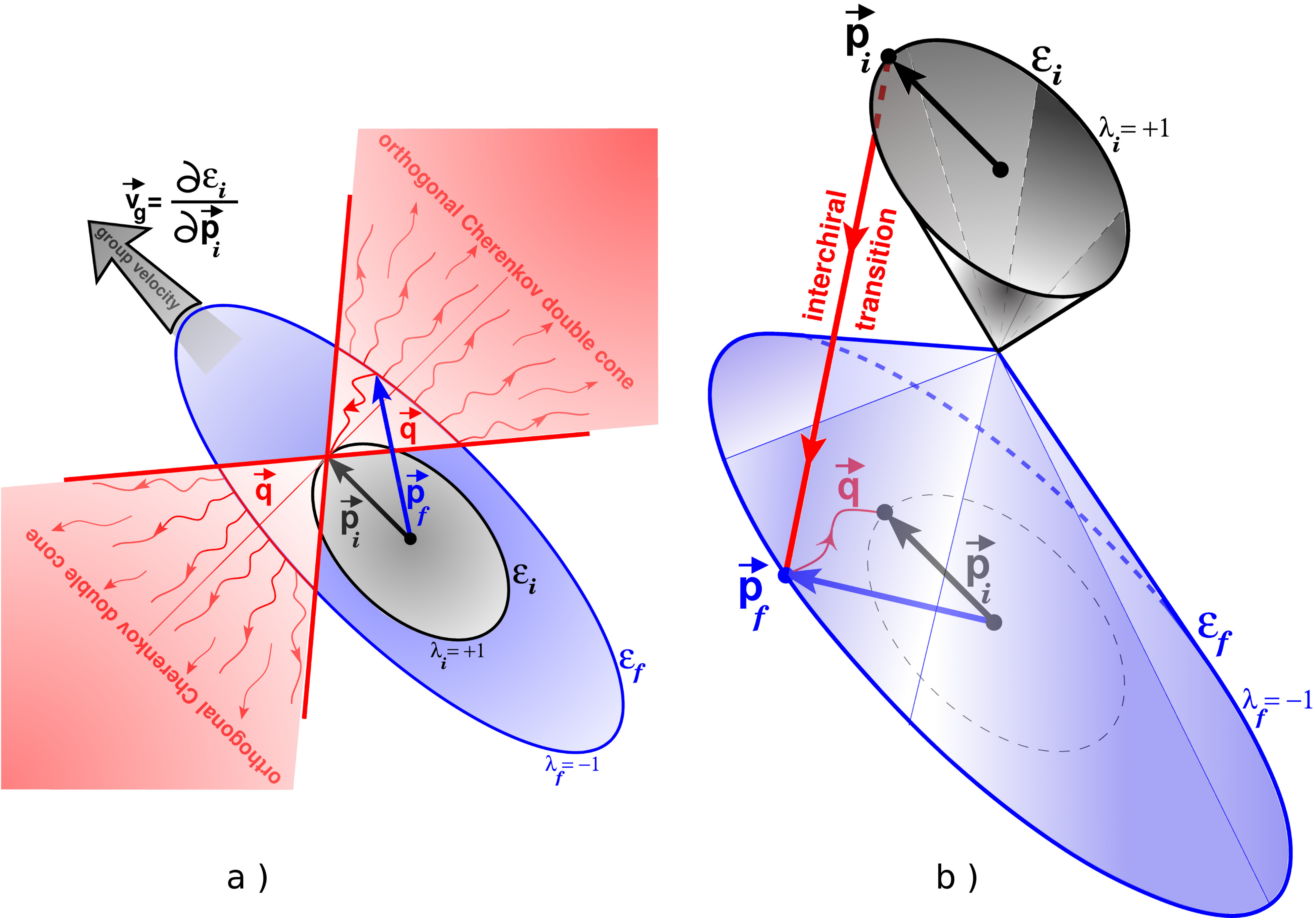}
\caption{\label{figure_1} A qualitative illustration of the orthogonal
  Cherenkov sound in a two-dimensional electron gas with Bychkov-Rashba and
  Dresselhaus spin-orbit interactions. Part a) shows the process of the
  orthogonal Cherenkov sound generation in the plane of the two-dimensional
  electron gas. Part b) illustrates the corresponding interchiral transition
  using a three-dimensional representation of the electron spectrum
  (Eqs. (\ref{SpEe}) and (\ref{AD})) at low energies and small momenta. A
  subsonic electron with the initial momentum $\vec{p}_i$ (black arrow) and
  chirality $\lambda_i=+1$ (the corresponding energy is
  $\varepsilon_i\equiv\varepsilon_{\lambda_i\vec{p}_i}$) changes its state via
  the electron-phonon interaction. The electron final state is characterized
  by the momentum $\vec{p}_f$ (blue arrow) and chirality $\lambda_f=-1$ (the
  corresponding energy is
  $\varepsilon_f\equiv\varepsilon_{\lambda_f\vec{p}_f}$). The electron group
  velocity in the initial state
  $\vec{v}_g=\partial\varepsilon_i/\partial\vec{p}_i$ (wide black arrow) has
  the same direction as the initial momentum $\vec{p}_i.$ The momentum and
  energy conservation laws admit the emission of a phonon (red wavy arrow
  connecting the vectors $\vec{p}_i$ and $\vec{p}_f$) within the double cone
  whose axis is orthogonal to $\vec{p}_i$ or $\vec{v}_g$. The phonon momentum
  and energy are $\vec{q}=\vec{p}_i-\vec{p}_f$,
  $\hbar\omega_\vec{q}=\varepsilon_i-\varepsilon_f$, respectively. Thus,
  instead of the normal forward or anomalous backward Cherenkov sound excited
  by supersonic particles within a cone whose axis is parallel to $\vec{p}_i$
  or $\vec{v}_g$, the subsonic particles in spin-orbit coupled systems may
  excite a unique Cherenkov sound within a double cone whose axis gets a
  quarter-turn with respect to $\vec{p}_i$ or $\vec{v}_g$, as shown by
  multiple phonons (multiple red wavy arrows) within the red area indicating
  the Cherenkov double cone.}
\end{figure}

A qualitative illustration of the orthogonal Cherenkov sound is shown in
Fig. \ref{figure_1}. In a two-dimensional electron gas with the Bychkov-Rashba
and Dresselhaus spin-orbit interactions the electronic states are
characterized (see below) by their chiralities $\lambda$ and momenta
$\vec{p}=(|\vec{p}|,\phi_\vec{p})$, where
$\cos\phi_\vec{p}\equiv p_x/|\vec{p}|$,
$\sin\phi_\vec{p}\equiv p_y/|\vec{p}|$. The energy spectrum consists of two
branches, the upper ($\lambda=+1$) and the lower ($\lambda=-1$) ones. The
isoenergy surfaces represent ellipses whose major axes is oriented along the
line specified by the beams with the polar angle  $3\pi/4$ or
$3\pi/4+\pi$. When the strengths of the two spin-orbit interaction mechanisms
are comparable the major axes of these ellipses get much longer than the minor
ones and the ellipses become extremely narrow. If an incident electron in the
upper branch has a momentum $\vec{p}_i$ (here and below the subscripts $i$ and
$f$ denote initial and final states, respectively) such that
$\phi_{\vec{p}_i}=3\pi/4$, then, being subsonic at low energies, it will be
unable to excite Cherenkov sound by virtue of intrachiral
($\lambda_i=+1\longrightarrow\lambda_f=+1$) transitions. However, interchiral
($\lambda_i=+1\longrightarrow\lambda_f=-1$) transitions with
$\phi_{\vec{p}_i-\vec{p}_f}\approx0,\pi$ will not generate Cherenkov sound
either because along these directions the electron energy changes very slowly
and, despite the additional energy due to the energy gap between states with
opposite chiralities, the electron energy loss is not compatible with its
momentum change to emit a phonon. On the other side, interchiral transitions
with $\phi_{\vec{p}_i-\vec{p}_f}\approx\pm\pi/2$ might generate Cherenkov
sound (red areas in Fig. \ref{figure_1}). Along these directions the electron
energy changes faster and in combination with the energy gap between states
with opposite chiralities such transitions may satisfy the energy and momentum
conservation laws, $\vec{p}_i=\vec{p}_f+\vec{q}$,
$\varepsilon_i=\varepsilon_f+\hbar\omega_\vec{q}$ ($\vec{q}$ is the phonon
momentum and $\hbar\omega_\vec{q}$ is its energy), see Fig. \ref{figure_1}.
\begin{figure}
\includegraphics[width=8.0 cm]{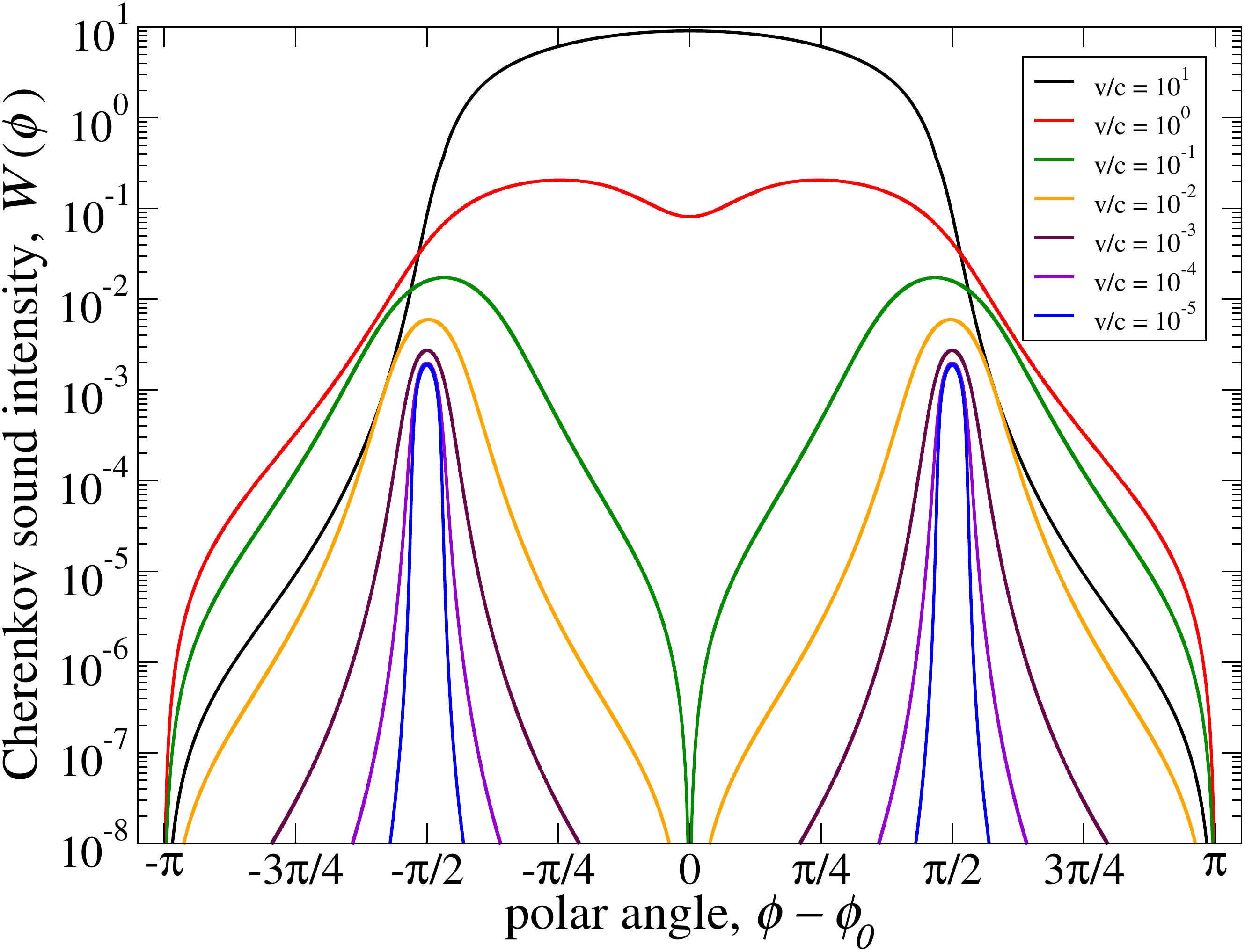}
\caption{\label{figure_2} Cherenkov sound intensity as a function of the polar
  angle for different values of $v/c$. The parameters are for the
  two-dimensional electron gas formed in an InAs quantum well structure with
  $c=4.2\cdot 10^3$ m/s, $m^\star=0.038\,m_0$ ($m_0$ is the free electron
  mass), $\alpha=0.15\cdot 10^{-11}$ eV$\cdot$m, $\beta/\alpha=0.85$. The
  sound is excited by electrons with $\lambda=+1$ and momenta with orientation
  $\phi_0=3\pi/4$. The electron group velocity,
  $\vec{v}_g\equiv\partial\varepsilon_{\lambda\vec{p}}/\partial\vec{p}$, has
  the same orientation.}
\end{figure}
\begin{figure}
\includegraphics[width=8.0 cm]{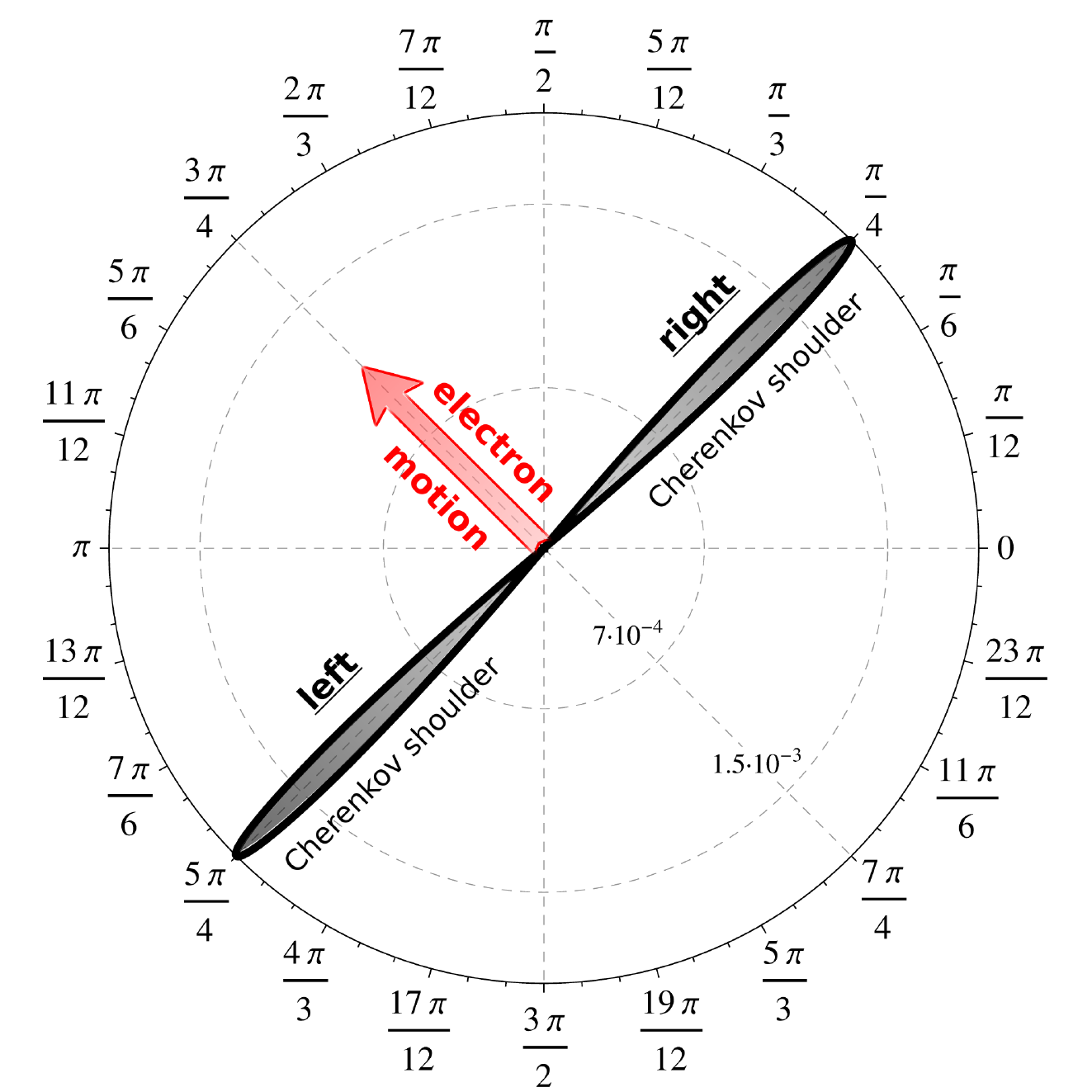}
\caption{\label{figure_3} The two-dimensional distribution of the Cherenkov
  sound in the plane of an InAs quantum well. Here $v/c=10^{-5}$, $c=4.2\cdot
  10^3$ m/s, $m^\star=0.038\,m_0$ ($m_0$ is the free electron mass),
  $\alpha=0.15\cdot 10^{-11}$ eV$\cdot$m, $\beta/\alpha=0.85$. The sound is
  excited by electrons with $\lambda=+1$ and momenta with orientation
  $\phi_0=3\pi/4$. The electron group velocity,
  $\vec{v}_g\equiv\partial\varepsilon_{\lambda\vec{p}}/\partial\vec{p}$, has
  the same orientation.}
\end{figure}

To quantitatively verify whether subsonic electrons may excite the Cherenkov
sound within the above mentioned qualitative scenario as well as to explore
its properties we briefly formulate below the computational scheme used to
obtain the Cherenkov sound intensity. The single-particle Hamiltonian is:
\begin{equation}
\hat{H}=\frac{\hat{\vec{p}}^2}{2m^\star}+
\frac{\alpha}{\hbar}(\hat{\sigma}_x\hat{p}_y-\hat{\sigma}_y\hat{p}_x)+
\frac{\beta}{\hbar}(\hat{\sigma}_x\hat{p}_x-\hat{\sigma}_y\hat{p}_y),
\label{SpHam}
\end{equation}
where $m^\star$ is the electron effective mass, $\alpha$ and $\beta$ are the
strengths of the Bychkov-Rashba and Dresselhaus spin-orbit interactions,
respectively. The eigenenergies of $\hat{H}$ are
\begin{equation}
\varepsilon_{\lambda\vec{p}}=\vec{p}^2/2m^\star+\lambda
p_{so}|\vec{p}|A_D(\vec{p})/m^\star,\quad p_{so}\equiv m^\star\alpha/\hbar,
\label{SpEe}
\end{equation}
\begin{equation}
A_D(\vec{p})\equiv\sqrt{1+(\beta/\alpha)^2+2(\beta/\alpha)\sin(2\phi_\vec{p})},
\label{AD}
\end{equation}
and its two component eigenspinors are
$(1/\sqrt{2})(1,\,-i\lambda\exp[i(\phi_\vec{p}-\tilde{\phi}_\vec{p})])^\text{T}$,
where
$\cos\tilde{\phi}_\vec{p}=[1+(\beta/\alpha)\sin(2\phi_\vec{p})]/A_D(\vec{p})$,
$\sin\tilde{\phi}_\vec{p}=-(\beta/\alpha)\cos(2\phi_\vec{p})/A_D(\vec{p})$. 

The phonon part \cite{AGD} of the total Hamiltonian is
\begin{equation}
\begin{split}
&\hat{H}_{ph}=\sum_{\vec{q}}\hbar\omega_\vec{q}(b^\dagger_\vec{q}b_\vec{q}+1/2)+g_{ph}\sum_\sigma\int d\vec{r}
\hat{\psi}^\dagger_\sigma(\vec{r})\hat{\psi}_\sigma(\vec{r})i\sum_{\vec{q}}\sqrt{\frac{\hbar\omega_\vec{q}}{2V}}
\biggl[\exp\biggl(i\frac{\vec{q}\vec{r}}{\hbar}\biggl)b_\vec{q}-\text{H.c.}\biggl],
\end{split}
\label{PhHam}
\end{equation}
where for the acoustic phonons $\hbar\omega_\vec{q}=c|\vec{q}|$ with $c$ being
the sound velocity. The first term in Eq. (\ref{PhHam}) describes the phonon
gas in terms of the second quantized operators
($[b_\vec{q},b^\dagger_{\vec{q}'}]=\delta_{\vec{q}\vec{q}'}$) while the second
one describes the electron-phonon interaction of strength $g_{ph}$ via the
coupling to the electron field operators
($\{\hat{\psi}_\sigma(\vec{r}),\hat{\psi}^\dagger_{\sigma'}(\vec{r}')\}=
\delta_{\sigma\sigma'}\delta(\vec{r}-\vec{r}')$).
\begin{figure}
\includegraphics[width=8.0 cm]{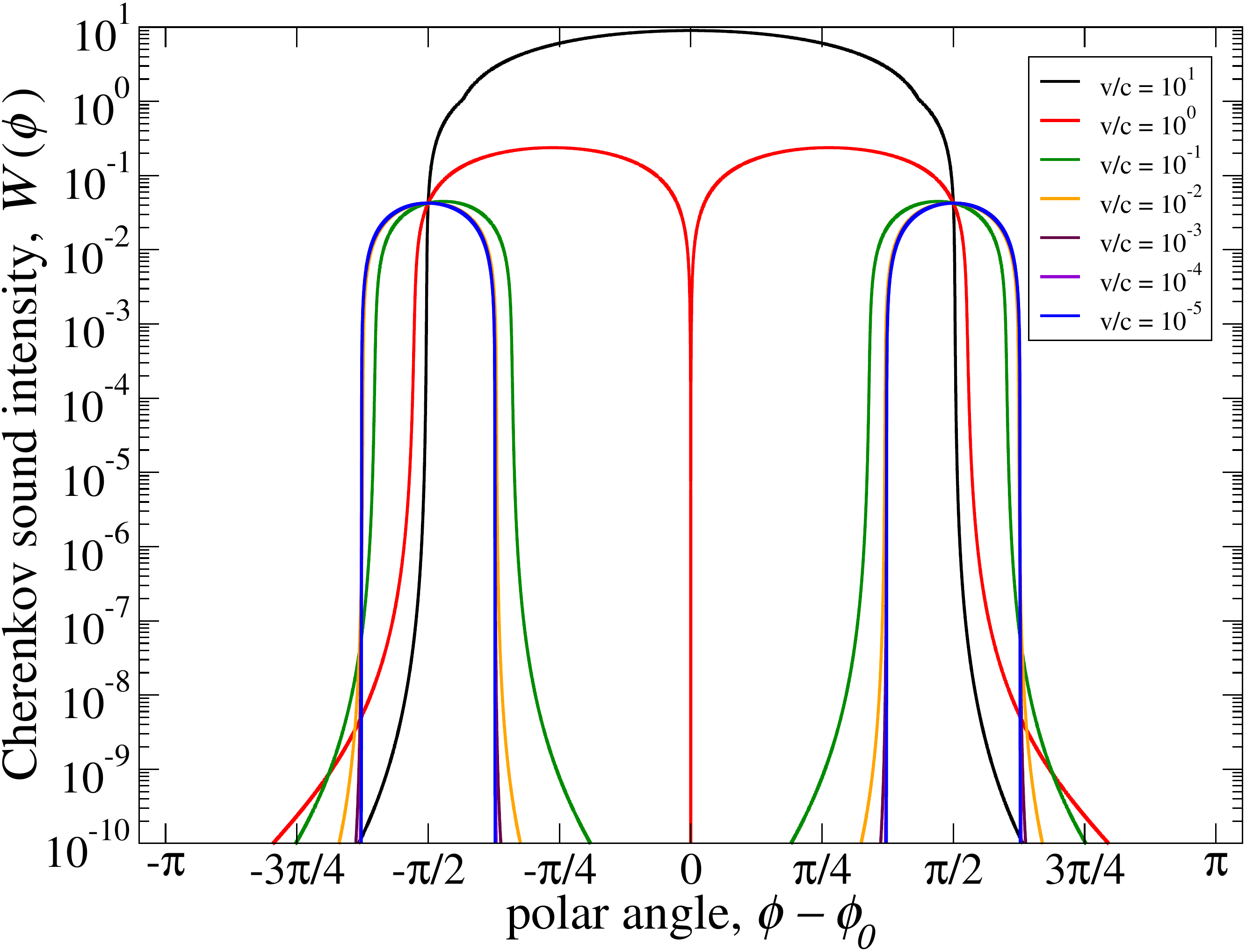}
\caption{\label{figure_4} Cherenkov sound intensity as a function of the polar
  angle for different values of $v/c$. The parameters are for the
  two-dimensional electron gas formed in an InAs quantum well structure with
  $c=4.2\cdot 10^3$ m/s, $m^\star=0.038\,m_0$ ($m_0$ is the free electron
  mass), $\alpha=0.15\cdot 10^{-11}$ eV$\cdot$m, $\beta=\alpha$. The sound is
  excited by electrons with $\lambda=+1$ and momenta with orientation
  $\phi_0=3\pi/4$. The electron group velocity,
  $\vec{v}_g\equiv\partial\varepsilon_{\lambda\vec{p}}/\partial\vec{p}$, has
  the same orientation.}
\end{figure}

Let us mention that the electron-phonon interaction in Eq. (\ref{PhHam}) is of
the deformation potential type. In noncentrosymmetric crystals the
piezoelectric electron-phonon coupling mechanism might also become important
and in principle for a quantitative analysis both electron-phonon interactions
should be considered. However, focusing on the qualitative aspect, we note
that the orthogonal Cherenkov sound results from the general conservation laws
of the energy and momentum and, therefore, its very existence can be predicted
within a minimal model taking into account for simplicity only one
electron-phonon interaction mechanism. It is our goal here to demonstrate
within a minimal model the existence of an unusual Cherenkov effect which, in
contrast to the known Cherenkov effects, is neither forward nor backward but,
to some extent, is of an intermediate orthogonal nature. Moreover, taking into
account possible applications in spintronic devices with a magnetic field
control the deformation potential electron-phonon coupling mechanism may
become dominant in InAs based structures \cite{Alcalde_2003}.

We follow the standard calculation
\cite{LS,Smirnov_2011,Smirnov_2013,Smirnov_2014} to obtain the self-energy
being the sum of the single-particle irreducible Feynman diagrams contributing
to the electron Green's function $G_{\lambda\vec{p}}(t-t')$ defined with
respect to the physical vacuum $|\Psi_0\rangle$,
$G_{\lambda\vec{p}}(t-t')=
-i\langle\Psi_0|T\psi_{\lambda\vec{p}}(t)\psi^\dagger_{\lambda\vec{p}}(t')|\Psi_0\rangle$,
where the electronic field operators are in the Heisenberg representation. We
calculate the self-energy in the second order in $g_{ph}$ which is equivalent
to the Fermi's golden rule but the self-energy approach has a more systematic
form easily generalized to derivations of the Cherenkov effect resulting from
higher orders in $g_{ph}$ as might be important in cases where the second
order Cherenkov effect is absent. The dimensionless Cherenkov sound intensity
as a function of the polar angle is obtained from the imaginary part of the
self-energy,
\begin{equation}
\text{Im}[\Sigma_{\lambda\vec{p}}(\varepsilon)]=
-\frac{g_{ph}^2m^{\star 2}c^2}{2\pi\hbar^3}\int_{-\pi}^{\pi}W(\phi)\,d\phi,
\label{SE}
\end{equation}
on the mass surface, $\varepsilon=\varepsilon_{\lambda\vec{p}}$, with
$\lambda=+1$. It is given as the sum of the two terms,
$W(\phi)=W_1(\phi)+W_2(\phi)$, originating, respectively, from intrachiral and
interchiral transitions:
\begin{equation}
W_{1,2}(\phi)=
\sum_j q'^2
\frac{1\pm\cos[\Phi(q',\phi)]}
{|h_{1,2}(q',\phi)|}\biggl|_{q'=q_j(\phi)},
\label{ChSndInt}
\end{equation}
where $q'\equiv|\vec{q}|/m^\star c$,
$h_{1,2}(q',\phi)\equiv(8/m^\star c^2)\partial_{q'}\psi_{1,2}(q',\phi)$,
$\psi_{1,2}(q',\phi)\equiv
\varepsilon_{\lambda=+1,\vec{p}_0}-
\varepsilon_{\lambda=\pm 1,\vec{p}_0-\vec{q}}-c|\vec{q}|$,
$\Phi(q',\phi)\equiv
\Delta\phi_{\vec{p}_0-\vec{q}}-\Delta\tilde{\phi}_{\vec{p}_0-\vec{q}}$,
$\Delta\phi_\vec{p}\equiv\phi_\vec{p}-\phi_0$,
$\Delta\tilde{\phi}_\vec{p}\equiv\tilde{\phi}_\vec{p}-\tilde{\phi}_0$,
$\phi_0\equiv\phi_{\vec{p}_0}$,
$\tilde{\phi}_0\equiv\tilde{\phi}_{\vec{p}_0}$,
$\vec{p}_0$ is the initial electron momentum and the summation is over all the
roots $q_j(\phi)$ of the momentum-energy conservation equations,
\begin{equation}
\begin{split}
&2\frac{v_{so}v}{c^2}A_D(\vec{p}_0)-q'^2+2q'\biggl(\frac{v}{c}\cos\phi-1\biggl)\mp\\
&\mp2\frac{v_{so}v}{c^2}A_D(\vec{p}_0-\vec{q})\sqrt{1+\biggl(\frac{cq'}{v}\biggl)^2-
2\frac{cq'}{v}\cos\phi}=0,
\end{split}
\label{MmEnConsv}
\end{equation}
where $v_{so}\equiv p_{so}/m^\star$ is the anomalous spin-dependent velocity
\cite{Adams_1959} and $v\equiv |\vec{p}_0|/m^\star$.

In Fig. \ref{figure_2} we show the total Cherenkov sound intensity,
$W(\phi)=W_1(\phi)+W_2(\phi)$. For large $v/c$ the Cherenkov sound is excited
by supersonic electrons. In this case the sound is generated by both
intrachiral and interchiral transitions and has the standard conic geometry
where the cone axis has the same orientation $\phi_0$ as the electron momentum
or group velocity and the cone angle extends up to $\pi$ as expected
\cite{Smirnov_2011} in spin-orbit coupled systems for supersonic
electrons. When $v/c$ decreases the intrachiral contribution to the Cherenkov
sound decays, the interchiral contribution starts to dominate and the sound
intensity rapidly decreases along the standard cone axis, {\it i.e.}, in the
vicinities of the angles $\phi=\phi_0$ and $\phi=\phi_0+\pi$. For smaller
$v/c$ these vicinities grow into broad areas which eventually enclasp the
whole space apart from the two enclaves around the two angles
$\phi=\phi_0\pm\pi/2$, {\it i.e.}, an orthogonal Cherenkov double cone forms
as the most elegantly demonstrated by the blue curve for which $v/c=10^{-5}$
and $|\vec{v}_g|/c\approx 0.0814$ (deep subsonic regime).

The actual distribution of the Cherenkov sound within the plane of the quantum
well is best visualized using the polar coordinates as is done in
Fig. \ref{figure_3}. The parameters of the quantum well are the same as the
ones used to obtain the data shown in Fig. \ref{figure_2} for $v/c=10^{-5}$
(deep subsonic regime, $|\vec{v}_g|/c\approx 0.0814$). In this representation
the orthogonal nature of the Cherenkov sound is clearly revealed: the sound
(dark areas) is localized mainly along the two directions orthogonal to the
electron momentum or group velocity direction (red arrow) and consists of the
left and right shoulders forming a narrow orthogonal Cherenkov double cone.

Finally, Fig. \ref{figure_4} demonstrates the total Cherenkov sound intensity
for the case $\alpha=\beta$ for the same values of $v/c$ as in
Fig. \ref{figure_2}. As one can see, here the Cherenkov sound has also an
orthogonal nature. The localization of the Cherenkov sound around the
direction perpendicular to $\vec{v}_g$ happens in this case faster because
already for $v/c=10^{-2}$ one reaches the deep subsonic regime,
$|\vec{v}_g|/c\approx 10^{-2}$. For even smaller values of $v/c$ all the sound
intensity curves collapse onto one curve representing the saturation 
limit corresponding to the linear dispersion of the initial electronic state
$\varepsilon_{\lambda=+1,\vec{p}}$ for small momenta, $|\vec{p}|\approx
0$. Note also that for $\alpha=\beta$ the size of the localization domain
also saturates so that the width of the peaks cannot be reduced by further
decreasing $v/c$.

In conclusion we would like to mention that, in addition to its unique
fundamental nature described above, the Cherenkov sound represents a general
and realistic channel of the electron energy dissipation. Usually one assumes
that at low energies the conventional Cherenkov dissipation is locked out
since all particles are subsonic. Our analysis, however, shows that this is
not the case in spin-orbit coupled systems and in practice one encounters the
problem of energy losses also in the subsonic regime which might be important
for phonon-limited low-field electron mobility \cite{Chen} being crucial for
efficient high clock frequency and low power spintronics. Another practical
aspect of the orthogonal Cherenkov sound is in diverse combinations of modern
spintronics and acoustics to build such devices as acoustic amplifiers
\cite{Komirenko_2000,Zhao_2009,Willatzen_2014} or sources of sound
\cite{Zhao_2013} based on the acoustic Cherenkov effect with a flexible
control of the sound direction determined by the electron spin dynamics in
contrast to the more conventional manipulation mechanisms based on the orbital
dynamics.

Support from the DFG under the program SFB 689 is acknowledged.

\section{Additional information}
{\bf Competing financial interests:} The author declares no competing
financial interests.

\end{document}